\documentstyle[aps,prl,multicol,epsfig,amstex]{revtex}

\title{Demixing in a single-peak distributed polydisperse
mixture of hard spheres}

\author{Jos\'e A.\ Cuesta\cite{email}}
\address{Grupo Interdisciplinar de Sistemas Complicados
(GISC), Departamento de Matem\'aticas, Escuela Polit\'ecnica
Superior, Universidad Carlos III de Madrid, c/ Butarque, 15, 28911 --
Legan\'es, Madrid, Spain}

\begin{document}
\draft
\preprint{MA/UC3M/11/1998}

\maketitle

\begin{abstract}
An analytic derivation of the spinodal of a polydisperse mixture
is presented. It holds for fluids whose excess free energy
can be accurately described by a function of a few
moments of the size distribution. It is shown that one such mixture
of hard spheres in the Percus-Yevick approximation
never demixes, despite its size distribution. In the
Boubl\'\i k-Mansoori-Carnahan-Starling-Leland approximation, though,
it demixes for a sufficiently wide log-normal size distribution.
The importance of this result is twofold: first, this distribution is
unimodal, and yet it phase separates; and second, log-normal
size distributions appear in many experimental contexts. The same
phenomenon is shown to occur for the fluid of parallel hard cubes.
\end{abstract}


\begin{multicols}{2}
\narrowtext

Our knowledge of the phase behavior of mixtures has increased
a lot in the last decade. Generally speaking, above a certain
concentration binary mixtures phase separate due to the
socalled depletion effect \cite{asakura}. This mechanism has
an entropic origin: mixtures phase separate when this produces
a sufficient gain in free volume to increase the entropy of the
system. More intuitively, depletion is an effective
attraction arising from the unbalance pressure resulting
when two solute particles are so close together
that no solvent particle fits in between them. Either way
we look upon it,
in real colloids this mechanism is supplemented by energetic
attractions or repulsions that enhance or inhibit the transition.
This is the reason why most theoretical research has focused in 
additive hard particle mixtures, in order
to ascertain to which extent depletion is the basic mechanism
of demixing

Binary hard spheres (HS), the simplest hard-particle mixture, has
proven very controversial in settling the question. More than
thirty years ago Lebowitz and Rowlinson \cite{lebowitz} showed
that according to the Percus-Yevick (PY) solution of the
Ornstein-Zernike
equation binary HS never demix; a few years later Vrij
\cite{vrij} extended this result to $p$-component HS mixtures
in the same approximation. However, the PY approximation has a
well known thermodynamical inconsistency, as it
yields two different equations of state, none of which being
very accurate at large densities. It is also well known
that a linear interpolation of both produces a rather accurate 
equation of state; in the case of mixtures, the latter is known
as the BMCSL equation of state \cite{boublik,mansoori}. In spite
of its higher accuracy, the authors showed that it predicts the
stability of any binary mixture.

The BMCSL equation of state is heuristic in its construction, so
this led Biben and Hansen \cite{biben} to consider the more accurate 
Rogers-Young closure approximation. In this way they found signs of a
spinodal instability for diameter ratios of the HS smaller
than 0.2. Several authors \cite{lekkerkerker,rosenfeld} subsequently
confirmed that other approximate
schemes yield the same instability, although for diameter ratios
strongly dependent on the approximation.
It was hence believed that phase separation does occur in sufficiently
asymmetric binary mixtures of HS, an idea further supported by
the demixing found in binary mixtures of parallel
hard cubes (PHC) both in computer simulations on a lattice
\cite{dijkstra} and from fundamental measure theory \cite{cuesta}
in continuum space.

Presently a different scheme is being accepted to describe demixing
of very asymmetric binary mixtures of hard particles \cite{note0}:
phase separation
occurs between a small-particle rich fluid and a large-particle
rich crystal \cite{imhof,poon,caccamo,noe,dijkstra2,yuri}, and
this coupling with the translational degrees of
freedom strongly enhances demixing. It may even cause, for very
large asymmetry, the appearance of an isostructural solid-solid
transition \cite{noe,dijkstra2,yuri}
as that found for narrow and deep attractive potentials---the reason
being that depletion induces one such potential between the large
particles \cite{noe,dijkstra2,yuri,biben2}.

But colloids are hardly mono or bidisperse. By their very nature,
colloidal particles are usually different from each other, 
and the preparation process of a colloidal suspension gives rise to
a size distribution of particles in the colloid. Hence colloids are
best modeled as polydisperse systems. Polydispersity has received
increasing attention in the past years
\cite{salacuse,gualteri,bolhuis,bartlett,sollich,warren,stell,sear}
because it causes strong qualitative effects on the phase behavior
of the monodisperse counterpart. Termination of the freezing transition
of a HS fluid \cite{bolhuis,bartlett}, regularization of adhesive 
potentials \cite{stell,yuri}, or appearance of a vapor-liquid 
transition in adhesive HS \cite{sear} are but a few examples.

Phase equilibrium in polydisperse systems is more complicated
than in monodisperse systems: equilibrium equations between coexisting
phases become functional equations when polydispersity is present
\cite{salacuse,gualteri}. Much of the effort has indeed concentrated
in taking advantage of extra symmetries (as the dependence of the excess
free energy on the moments of the distribution \cite{sollich,warren}),
or in mapping the polydisperse system into a simpler one (e.g.\
a binary mixture \cite{bartlett}), with the purpose of reducing
equilibrium to a few algebraic equations. Specifically, by means of
a moment-based formalism \cite{sollich,warren}
Warren \cite{warren2} has shown that while a binary fluid of HS
in the BMCSL approximation never demixes, it does if enough
polydispersity is introduced and the diameter ratio of the
two ``main'' species is sufficiently small. This unexpected result
proves that the effect of polydispersity can be more subtle and 
nontrivial than one can tell a priori.

\begin{figure}
\epsfig{file=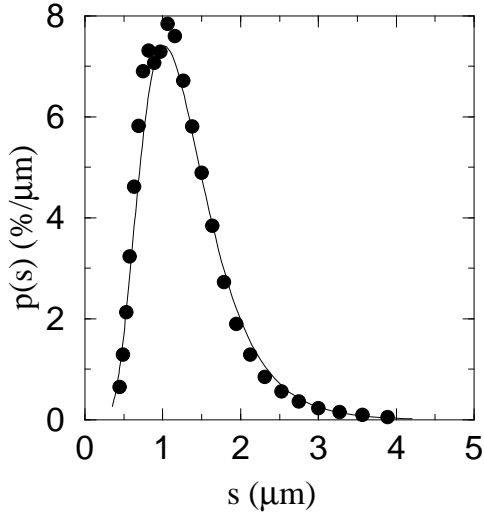, width=2.7in}
\caption[]{Distribution, $p(s)$ (in percentage per $\mu$m),
of diameters, $s$ (in $\mu$m), for a suspension of {\em mullite}
(3Al$_2$O$_3\cdot 2$SiO$_2$), a ceramic powder used in structural
ceramic applications. Bullets are the experimental data; the line
is a fit of a log-normal distribution of mean $\overline{x}=1.57$
$\mu$m and variance $w=0.43\overline{x}$. (Experimental data courtesy
of R.\ Moreno, Instituto de Cer\'amica y Vidrio, CSIC, Madrid, Spain.)}
\label{experim}
\end{figure}

In this letter I will push this result much further. Though 
polydisperse, Warren's system has two clearly differentiated
species. I will show that for demixing to occur this is not
necessary if the size distribution decays sufficiently slowly. Again
using a 
moment-based formalism (equivalent to Warren's but based on different
grounds) to obtain a spinodal of an arbitrary mixture (with several
species, polydisperse, or both), I will prove that BCSML polydisperse
HS with a {\em log-normal} size distribution (i.e.\ the logarithm
of the diameters follows a normal distribution) phase separate provided
the variance of the distribution is above a threshold. This result
is surprising because one such distribution is unimodal. It is thus
clear that the
long-size tail plays a crucial role in inducing demixing, but although
slowly decaying, this tail decays faster than any power-law. In
addition, it is important to stress that,
contrary to what one might believe, log-normal distributions are
very common in many polydisperse systems: ultrafine metal particles,
paint and rubber pigments, photographic emulsions, dust particles,
aerosols, cloud droplets, or suspensions of ceramic powders 
(see \cite{granqvist} and references therein; see also
Fig.\ \ref{experim}), etc.
Therefore this effect should be experimentally accessible,
provided any of these systems can be prepared
with sufficiently large size dispersion.

In what follows I will briefly describe the formalism to obtain the
spinodal in the case that the excess free energy of a mixture depends on
a few moments of the size distribution. Then I will apply this formalism
to polydisperse HS and PHC in scaled particle approximation (equivalent
to PY in the case of HS), and I will finally prove the above mentioned
result.

Suppose we have a multicomponent hard particle mixture, whose free 
energy per unit volume is $f\equiv \beta F/V=f^{\rm id}+f^{\rm ex}$,
with $f^{\rm id}=\sum_i\rho_i[\log{\cal V}_i\rho_i-1]$, ${\cal V}_i$
the thermal volume of species $i$, $\beta$ the reciprocal temperature
(in units of the Boltzmann constant), and $\rho_i$ the number
density of species $i$. Let us further
assume that $f^{\rm ex}=\phi(\xi)$, $\xi$ denoting the set of moments
$\{\xi_0,\xi_1,\dots,\xi_p\}$, where $\xi_k=\sum_i\sigma_i^k\rho_i$
($\sigma_i$ is the diameter of species $i$).
Stability of the mixture requires $f$ to be a convex function in all
of its variables. As in any hard particle system $f$ depends on
temperature through an additive term (the free energy of the ideal
gas) and hence it is always convex with respect to this variable. 
The study of stability is thus restricted to the set of densities of
the species involved. If we define the matrix
\begin{equation}
{\sf M}\equiv\frac{\partial^2 f}{\partial\rho_i\partial\rho_j}=
\frac{1}{\rho_i}\delta_{ij}+\frac{\partial^2\phi}
{\partial\rho_i\partial\rho_j},  
\end{equation}
then stability requires it to be positive definite. As it is so for
low densities (ideal mixtures are always stable) the spinodal is
usually described as the set of points in the densities 
space where $|{\sf M}|=0$ ($|\cdot|$ denotes the determinant)
\cite{lebowitz,callen}. This condition is equivalent to ${\sf M}$
having a zero eigenvalue, i.e.\ ${\sf M}\cdot{\bf u}=0$ for some
vector ${\bf u}\ne 0$. Let us define for convenience a new vector
${\bf e}$ as $u_i=\rho_ie_i$; then ${\sf M}\cdot{\bf u}=0$ means
\begin{equation}
e_i=-\sum_j\frac{\partial^2\phi}{\partial\rho_i\partial\rho_j}
\rho_je_j  .
\label{eq1}
\end{equation}

For a polydisperse mixture, if $\rho(s)=\rho p(s)$, where $\rho$ is
the total number density of particles and $p(s)$ is the size
probability distribution (in terms of a dimensionless diameter $s$,
for instance), $f^{\rm id}=\int ds\,\rho(s)[\log{\cal V}(s)
\rho(s)-1]$ \cite{salacuse}, $\xi_k=\int ds\,s^k\rho(s)$,
but $f^{\rm ex}$ is still assumed to be
of the form $f^{\rm ex}=\phi(\xi)$. The stability condition is now
the positive definiteness of the integral operator whose kernel is
defined by
\begin{equation}
{\cal M}(s,t)\equiv\frac{\delta^2 f}{\delta\rho(s)\delta\rho(t)}=
\frac{1}{\rho(s)}\delta(s-t)+\frac{\delta^2\phi}{\delta\rho(s)
\delta\rho(t)} \, .
\end{equation}
While the determinant condition has no direct equivalent for an 
integral operator, the zero-eigenvalue one is a straightforward 
extension of (\ref{eq1}):
\begin{equation}
e(s)=-\int dt\,\frac{\delta^2\phi}{\delta\rho(s)\delta\rho(t)}
\rho(t)e(t) \, .
\label{eq2}
\end{equation}
But we can simplify this equation by using
\begin{equation}
\frac{\delta^2\phi}{\delta\rho(s)\delta\rho(t)}=
\sum_{k,l=0}^p\frac{\partial^2\phi}{\partial\xi_k\partial\xi_l}
s^kt^l,
\end{equation}
which transforms Eq.\ (\ref{eq2}) into
\begin{equation}
e(s)=-\sum_{k,l=0}^p\Phi_{kl}s^k\alpha_l , \qquad
\Phi_{kl}\equiv\frac{\partial^2\phi}{\partial\xi_k\partial\xi_l}, 
\label{eq3}
\end{equation}
where $\alpha_l=\int ds\,s^l\rho(s)e(s)$. Substitution of
Eq.\ (\ref{eq3}) into the latter definition yields \cite{note}
\begin{equation}
\alpha_n=-\sum_{k,l=0}^p\xi_{n+k}\Phi_{kl}\alpha_l . 
\end{equation}
For this equation to have nonzero solutions for $\alpha$ we must 
have $|{\sf Q}|=0$, with
\begin{equation}
{\sf Q}_{mn}\equiv\delta_{mn}+\sum_{k=0}^p\xi_{m+k}\Phi_{kn} ,
\quad m,n=0,\dots,p .
\end{equation}

For convenience we introduce the variables
$y_k\equiv\xi_k/(1-\xi_3)$
and define two new matrices, ${\sf Y}$ and
$\Omega$, by ${\sf Y}_{mn}\equiv y_{m+n}$ and $\Omega\equiv
(1-\xi_3)\Phi_{mn}$. Then the condition reads ${\cal D}\equiv
|{\Bbb I}+{\sf Y}\cdot\Omega|=0$, with ${\Bbb I}$ the identity matrix.

Many scaled-particle theories can be written
\cite{cuesta,cuesta2,rosenfeld2}
\begin{equation}
\phi=-\xi_0\ln(1-\xi_3)+A\frac{\xi_1\xi_2}{1-\xi_3}+
B\frac{\xi_2^3}{(1-\xi_3)^2}.
\label{SPT}
\end{equation}
(Notice that if $\phi$ and $\xi_k$ are all multiplied by the same
constant the product ${\sf Y}\cdot\Omega$ remains invariant, so we
will assume these variables defined up to a constant.)
For this particular choice of $\phi$ we will have
\begin{equation}
\Omega=\left(\begin{array}{cccc}
0 & 0 & 0 & 1 \\
0 & 0 & A & Ay_2 \\
0 & A & 6By_2 & Ay_1+6By_2^2 \\
1 & Ay_2 & Ay_1+6By_2^2 & y_0+2Ay_1y_2+6By_3^3
\end{array}\right). \label{matrixOm}
\end{equation}
It is then straightforward to check that 
\begin{equation}
{\cal D}=\frac{[1+(1-A)\xi_3]^2+(6B-A^2)\xi_2\xi_4}{(1-\xi_3)^4},
\end{equation}
where we have replaced the $y$'s in terms of the $\xi$'s;
$\xi_3$ is the volume fraction of the fluid. We are now ready to
discuss specific cases.

Percus-Yevick HS correspond to choosing
$A=3$, $B=3/2$ (definitions of $\phi$
and $\xi_k$ carry an extra $\pi/6$ factor); then ${\cal D}=
(1+2\xi_3)^2/(1-\xi_3)^4$, the result obtained by Lebowitz and Rowlinson
\cite{lebowitz} for the binary mixture and later generalised by Vrij
\cite{vrij} for a multicomponent mixture. The validity of this result
has now been extended for a polydisperse mixture \cite{note}.

Scaled-particle free energy for PHC is given by \cite{cuesta,cuesta2}
$A=3$, $B=1$. Then 
\begin{equation}
{\cal D}=[(1+2\xi_3)^2-3\xi_2\xi_4]/(1-\xi_3)^4 .
\end{equation}
By a suitable choice of $\xi_2$ and $\xi_4$ we can make ${\cal D}=0$.
If we consider a binary mixture we recover the result of Ref.\ 
\onlinecite{cuesta}. By defining $m_k\equiv\xi_k/\xi_3=\int ds\,s^kp(s)$,
the mixture of PHC in this approximation will be stable provided
$K(\xi_3) > m_2m_4$, where $K(\xi_3)\equiv(1+2\xi_3)^2/(3\xi_3^2)$.
The function $K(x)$ diverges at $x=0$ and it monotonically decreases
to 3 as $\xi_3$ approaches 1. Then $m_2m_4>3$ is the condition to find
demixing, and the larger the product $m_2m_4$ the smaller the 
value of $\xi_3$ at which it appears.

Schulz distribution is a common choice in studying polydispersity
\cite{salacuse}. If the mean is set to 1, it can be written as
\begin{equation}
p(s)=w^{-2}\Gamma(w^{-2})(w^{-2}s)^{w^{-2}-1}e^{-w^{-2}s},
\end{equation}
with $0<w<1$ the variance. Its moments are
$\xi_n=\rho\prod_{k=0}^{n-1}(1+kw^2)$ for $n\geq 1$, so
$m_2m_4=(1+3w^2)/(1+2w^2)<4/3$.
Therefore there is no demixing for such a distribution. It is easy
to check that there is no demixing either for a uniform distribution
in any interval of diameters. However if we consider a log-normal
distribution with mean 1 and variance $0<w<\infty$,
\begin{equation}
p(s)=\frac{1+w^2}{\sqrt{2\pi\ln(1+w^2)}}\exp\left\{
-\frac{\ln^2[(1+w^2)^{3/2}s]}{2\ln(1+w^2)} \right\} ,
\end{equation}
whose moments are given by $\xi_n=\rho(1+w^2)^{n(n-1)/2}$, then
$m_2m_4=1+w^2$,
which, by increasing $w$, can be made arbitrarily large. There is
thus demixing for a log-normal distribution provided $w>\sqrt{2}$.

The latter result rises the question whether the same occurs for
BMCSL HS. In this case the free energy is given by \cite{salacuse}
\begin{equation}
\phi= \left(\frac{\xi_2^3}{\xi_3^2}-\xi_0\right)\ln(1-\xi_3)+
\frac{3\xi_1\xi_2}{1-\xi_3}+\frac{\xi_2^3}{\xi_3(1-\xi_3)^2} ,
\end{equation}
where, as in PY, $\phi$ and $\xi_k$ are defined with an extra $\pi/6$
factor. The difference with respect to the PY one is 
\begin{equation}
\Delta\phi=\xi_2^3\left\{
\frac{1}{\xi_3^2}\ln(1-\xi_3)+\frac{1-3\xi_3/2}{\xi_3(1-\xi_3)^2}
\right\} ,
\end{equation}
where $\Delta\phi\equiv\phi_{\rm BMCSL}-\phi_{\rm PY}$.
The corresponding matrix $\Omega_{\rm BMCSL}=\Omega_{\rm PY}+
\Delta\Omega$, where $\Delta\Omega$ is obtained from the second
derivatives of $\Delta\phi$ (its only nonzero elements 
$\Delta\Omega_{jk}$ are those with $j,k=2,3$).

\begin{figure}
\epsfig{file=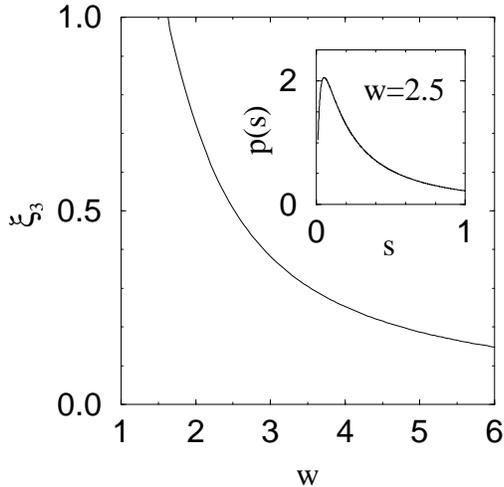, width=2.7in}
\caption[]{Volume fraction, $\xi_3$, as a function of the variance of
the log-normal size distribution, $w$, at which a spinodal instability
appears. Inset: size distribution for the particular value $w=2.5$.}
\label{spinodal}
\end{figure}

The resulting expression for ${\cal D}$ with the moments replaced by
those of a log-normal distribution is a complicated formula relating
the volume fraction, $\xi_3$, and the variance, $w$; nevertheless,
the spinodal can be numerically determined (Fig.\ \ref{spinodal}).
It can be seen that the threshold variance for demixing
is $w\approx 1.6$. The inset shows the size distribution
for a variance for which the spinodal occurs at liquid densities,
in order to illustrate the long tail at large particle sizes. Comparing
with Fig.\ \ref{experim} and with the figures of Ref.\ 
\onlinecite{granqvist} we can see that standard experimental
samples are still far from reaching the spinodal.

In summary, I have introduced a simple formalism to compute the
spinodal of fluids whose excess free energy can be 
described as a function of a few moments of the size distribution
of a mixture. By means of this formalism I have proven that a
polydisperse mixture of PY HS never demixes despite its size distribution.
I have also shown that a polydisperse mixture of PHC may demix only
if the size distribution is long-tailed at large sizes, as for
instance a log-normal distribution. More importantly,
polydisperse BMCSL HS, also log-normally distributed, do demix for a
sufficiently large variance. Remarkably, log-normal
distributions have a single characteristic size, so the driving
mechanism of the transition in these systems is not as clearcut
as in the case of binary mixtures. Finally, sizes in
many polydisperse systems are log-normally distributed.
Present experimentally available systems, though, are far from
the spinodal as predicted by this work. I hope that this result
encourages experimental work to achieve this limit and verify
the conclusions reported here.

It is a pleasure to thank R.\ P.\ Sear and P.\ Warren for
useful discussions. I am also grateful to A.\ Garc\'\i a,
R.\ Cuerno, R.\ Moreno, and A.\ S\'anchez for their invaluable
help. This work is supported by the
Direcci\'on General de Ense\~nanza Superior (Spain),
project no.\ PB96-0119.

\end{multicols}


\vspace*{-2mm}
\begin{references}
\bibitem[*]{email} E-mail: {\tt cuesta@@math.uc3m.es}
\bibitem{asakura} S.\ Asakura and F.\ Oosawa, \jcp {\bf 22}, 1255
    (1954).
\bibitem{lebowitz} J.\ L.\ Lebowitz and J.\ S.\ Rowlinson, \jcp {\bf 41},
    133 (1964).
\bibitem{vrij} A.\ Vrij, \jcp {\bf 69}, 1742 (1978).
\bibitem{boublik} T.\ Boublik, \jcp {\bf 53}, 471 (1970).
\bibitem{mansoori} G.\ A.\ Mansoori, N.\ F.\ Carnahan, K.\ E.\ Starling,
    and T.\ W.\ Leland, Jr., \jcp {\bf 77}, 3714 (1982).
\bibitem{biben} T.\ Biben and J.-P.\ Hansen, \prl {\bf 66}, 2215 (1991).
\bibitem{lekkerkerker} H.\ N.\ W.\ Lekkerkerker and A.\ Stroobants, Physica
    A {\bf 195}, 387 (1993).
\bibitem{rosenfeld} Y.\ Rosenfeld, \prl {\bf 72}, 3831 (1994).
\bibitem{dijkstra} M.\ Dijkstra and D.\ Frenkel, \prl {\bf 72}, 298 (1994);
    M.\ Dijkstra, D.\ Frenkel and J.-P.\ Hansen, \jcp {\bf 101}, 3179
    (1994).
\bibitem{cuesta} J.\ A.\ Cuesta, \prl {\bf 76}, 3742 (1996).
\bibitem{note0} This general scheme may change if the particles have
    internal degrees of freedom (e.g.\ rotations); see M.\ Dijkstra and
    R.\ van Roij, \pre {\bf 56}, 5594 (1997).
\bibitem{imhof} A.\ Imhof and J.\ K.\ G.\ Dhont, \prl {\bf 75}, 1662 (1995).
\bibitem{poon} W.\ C.\ K.\ Poon and P.\ B.\ Warren, Europhys.\ Lett.\
    {\bf 28}, 513 (1994).
\bibitem{caccamo} C.\ Caccamo and G.\ Pellicane, Physica A {\bf 235},
    149 (1997).
\bibitem{noe} N.\ Garc\'\i a-Almarza and E.\ Enciso in {\em Proceedings
    of the VIII Spanish Meeting on Statistical Physics FISES '97},
    p.\ 159, J.\ A.\ Cuesta and A.\ S\'anchez, eds.\ (Editorial del
    CIEMAT, Madrid, 1998).
\bibitem{dijkstra2} M.\ Dijkstra, R.\ van Roij, and R.\ Evans, submitted
    to \prl (1998).
\bibitem{yuri} Y.\ Mart\'\i nez--Rat\'on and J.\ A.\ Cuesta, preprint
    cond-mat/9804225 (1998).
\bibitem{biben2} T.\ Biben, P.\ Bladon, and D.\ Frenkel, J.\ Phys.:
    Condens.\ Matter {\bf 8}, 10799 (1996).
\bibitem{salacuse} J.\ J.\ Salacuse and G.\ Stell, \jcp {\bf 77}, 3714
    (1982).
\bibitem{gualteri} J.\ A.\ Gualteri, J.\ M.\ Kincaid, and G.\ Morrison,
    \jcp {\bf 77}, 521 (1982).
\bibitem{bolhuis} P.\ G.\ Bolhuis and D.\ A.\ Kofke, \pre {\bf 54},
    634 (1996).
\bibitem{bartlett} P.\ Bartlett, \jcp {\bf 107}, 188 (1997).
\bibitem{sollich} P.\ Sollich and M.\ E.\ Cates, \prl {\bf 80}, 1365
    (1998).
\bibitem{warren} P.\ B.\ Warren, \prl {\bf 80}, 1369 (1998).
\bibitem{stell} G.\ Stell, J.\ Stat.\ Phys.\ {\bf 63}, 1203 (1991).
\bibitem{sear} R.~P.~Sear, preprint cond-mat/9805201 (1998).
\bibitem{warren2} P.\ B.\ Warren, preprint (1998).
\bibitem{granqvist} C.\ G.\ Granqvist and R.\ A.\ Buhrman, Appl.\
    Phys.\ Lett.\ {\bf 27}, 693 (1975); J.\ Appl.\ Phys.\ {\bf 47}, 2200
    (1976).
\bibitem{callen} H.\ B.\ Callen, {\em Thermodynamics} (Wiley, N.\ Y.,
    1960).
\bibitem{cuesta2} J.\ A.\ Cuesta and Y.\ Mart\'\i nez-Rat\'on, \prl
    {\bf 78}, 3681 (1997); \jcp {\bf 107}, 6379 (1997).
\bibitem{note} This result can also be obtained, by the same procedure,
    from Eq.\ (\ref{eq1}), and so is valid for both multicomponent and
    polydisperse mixtures (or a combination of both).
\bibitem{rosenfeld2} Y.\ Rosenfeld, \pre {\bf 50}, R3318 (1994).
\end{references}
\end{document}